# Bilayer Isotropic Thermal Cloak

Tiancheng Han[1*], Xue Bai[1,2*], John T. L. Thong[1], Baowen Li[2], and Cheng-Wei Qiu[1#]

[1]Department of Electrical and Computer Engineering, National University of Singapore, Kent Ridge 119620, Republic of Singapore.    [#]E-mail: eleqc@nus.edu.sg

[2]Department of Physics and Centre for Computational Science and Engineering, National University of Singapore, Singapore 117546, Republic of Singapore.

**Abstract:** Invisibility has attracted intensive research in various communities, e.g., optics, electromagnetics, acoustics, thermodynamics, etc. However, the most of them have only been experimentally achieved by virtue of simplified approaches due to their inhomogeneous and extreme parameters imposed by transformation-optic method, and usually require challenging realization with metamaterials. In this paper, we demonstrate an advanced bilayer thermal cloak with naturally available materials first time. This scheme, directly from thermal conduction equation, has been validated as an exact cloak rather than a reduced one, and we experimentally confirmed its perfect performance (heat-front maintenance and heat protection) in an actual setup. The proposed scheme may open a new avenue to control the diffusive heat flow in ways inconceivable with phonons.

## Introduction

Rendering an object invisible has been a long-standing dream for many researchers over the decades. Recently, many significant achievements of invisibility cloaking have been motivated thanks to the pioneering theoretical works[1-4]. However, most cloak realization usually requires extreme constitutive parameters (inhomogeneous, anisotropic, and even singular). For the sake of feasibility, several reduced cloaks have been experimentally demonstrated at the price of giving up exactness[5-8]. Another strategy to further reduce a full cloak into something more practical is the carpet cloak[9], with several experimental realizations[10-15], which attempt to hide an object on the ground and mimic a half-infinite vacuum space. More recently, calcite crystals are employed to fabricate carpet cloaks based on linear transfroamtion[16,17], and one dimensional full-parameter cloak constructed with metamaterials have been experimentally demonstrated[18]. Readers may find more information in recent review papers[19,20]. Decoupling electric and magnetic effects, static magnetic cloak[21,22] and static electric cloak[23] have been experimentally realized using ferromagnetic-superconductor materials and resistor networks, respectively. In addition to manipulation of electromagnetic wave[5-23], the theoretical tool of coordinate transformation has been extended to acoustic waves[24,25] and heat flux[26-29].

On the basis of the invariance of heat conduction equation under coordinate transformation, transformation thermodynamics has provided a new method to manipulate heat flux at will[26]. Through tailoring inhomogeneity and anisotropy of conductivities (as well as specific heat and material density), transient thermal cloaking has been experimentally demonstrated recently[27,28]. In addition, steady-state thermal cloaking can be designed with only anisotropic conductivities and its construction can be further simplified by utilizing the multilayered composite approach as demonstrated both experimentally[29] and theoretically[30]. However, previous thermal cloaking[27-30] are usually anisotropic and (or) inhomogeneous, and needs to be reduced for experimental realization at the price of giving up exactness. Therefore, the practical applications of thermal cloak may be limited. Surprisingly, an exact magnetostatic cloak that has been recently realized using a layer of ferromagnetic material and a layer of superconductor, though it works at much lower temperature (77 K)[21,22]. This motivates us to explore an exact thermal cloak with extremely simple structure, only employing naturally available materials.

Here, we experimentally first time demonstrate a bilayer thermal cloak with naturally available materials, which has been theoretically validated as an exact cloak from first principles (not a reduced cloak). The contribution of this study is threefold. First, the bilayer cloak does not rely on transformation optics[3,4], thus can avoid the problems present in previous cloaking proposals, such as extreme parameters (inhomogeneous, anisotropic, and singular) and complicated fabrication[27-29]. Second, extremely simple bilayer structure demonstrates nearly perfect performance only employing naturally available materials, which indicates that the proposed scheme may be readily for engineering application. Third, the bilayer cloak is carefully investigated both in steady state and time-dependent case, which presents the excellent performance on heat-front maintenance and heat protection.

**Theoretical Analysis**

The bilayer thermal cloak is schematically illustrated in Fig. 1 (a), which is composed of an inner layer ($a<r<b$) and an outer layer ($b<r<c$) with conductivity of $\kappa_2$ and $\kappa_3$, respectively. The conductivity of the background is $\kappa_b$. We start from the general three-dimensional (3D) theoretical analysis, directly form thermal conduction equation. Heat flows spontaneously from a high temperature region toward a low temperature region. For a steady state and without heat source, the temperature satisfies $\nabla \cdot (\kappa \nabla T) = 0$ in all the regions of space, they can be generally expressed as

$$T_i = \sum_{m=1}^{\infty} \left[ A_m^i r^m + B_m^i r^{-m-1} \right] P_m(\cos\theta) \tag{1}$$

where $A_m^i$ and $B_m^i$ ($i$=1, 2, 3, 4) are constants to be determined by the boundary conditions and $T_i$ denotes the temperature in different regions: $i$=1 for the cloaking region ($r<a$), $i$=2 for the inner layer ($a<r<b$), $i$=3 for the outer layer ($b<r<c$), and $i$=4 for the exterior region ($r>c$).

A linear temperature distribution with uniform temperature gradient $t_0$ is externally applied in the z direction. Taking into account that $T_4$ should tend to $-t_0 r \cos\theta$ when $r \to \infty$, we only need to consider $m$=1. Because $T_1$ is limited when $r \to 0$, we can obtain $B_1^1 = 0$. Owing to the temperature potential and the normal component of heat flux vector being continuous across the interfaces, we have

$$\begin{cases} T_i \big|_{r=a,b,c} = T_{i+1} \big|_{r=a,b,c} \\ \kappa_i \dfrac{\partial T_i}{\partial r} \bigg|_{r=a,b,c} = \kappa_{i+1} \dfrac{\partial T_{i+1}}{\partial r} \bigg|_{r=a,b,c} \end{cases} \tag{2}$$

Here, $\kappa_4 = \kappa_b$ and $\kappa_1$ is the thermal conductivity of the cloaking object. Considering that the inner layer is perfect insulation material, i.e. $\kappa_2 = 0$, this ensures that an external field does not penetrate inside the cloaking region and the only task is to eliminate the external-field distortion. By substituting Eq. (1) into Eq. (2), we can obtain

$$B_1^4 = t_0 c^3 \frac{\kappa_3(2c^3 - 2b^3) - \kappa_b(2c^3 + b^3)}{\kappa_3(2c^3 - 2b^3) + 2\kappa_b(2c^3 + b^3)} \tag{3}$$

By setting $B_1^4 = 0$, we can obtain

$$\kappa_3 = \frac{2c^3 + b^3}{2(c^3 - b^3)} \kappa_b \tag{4}$$

Obviously, an ideal bilayer thermal cloak may be achieved as long as Eq. (4) is fulfilled. Eq. (4) represents that the third one can be uniquely determined if arbitrary two of $\kappa_3$, $\kappa_b$, $c/b$ are known. If $\kappa_3$ and $\kappa_b$ are given, $c$ is proportional to $b$, which means that the geometrical size of the cloak can be arbitrarily tuned without changing the materials of the bilayer cloak.

When we consider two-dimensional (2D) bilayer cloak in Fig. 1(b), analysis analogy to 3D case may obtain the relationship expressed as

$$\kappa_3 = \frac{c^2 + b^2}{c^2 - b^2} \kappa_b \tag{5}$$

## Simulation Results

The performance of the proposed bilayer cloak is numerically verified based on finite element method (FEM), in which the materials of the bilayer cloak are chosen to be the same as those in later experiments: inner layer, outer layer, and background material are polystyrene, alloy, and sealant with conductivity of 0.03 W/mK, 9.8 W/mK, and 2.3 W/mK, respectively. Based on the exact theoretical analysis, it is known that the inner layer of an ideal bilayer cloak should be filled with insulation material, i.e. $\kappa_2 \to 0$. However, ideal insulation material can not be obtained in practice. Here, polystyrene is utilized as insulation material whose conductivity is about 1/77 compared to the background material. This is exactly a nearly perfect approximation of ideal bilayer cloak, which will be conformed by the following simulation and measurement results.

For a 3D bilayer cloak, we choose $a$=6 mm, $b$=9 mm, and $c$=10.2 mm. An aluminum sphere with conductivity of 205 W/mK is placed in the cloaking region. Fig. 2 shows the simulated temperature distributions, in which (a) and (d) illustrate the perturbation (aluminum sphere) without and with the bilayer cloak, respectively. As expected, the central region without cloak makes the isothermal surfaces and thermal flux lines significantly distorted, thus leading to the object visible (detected). However, when the central region is wrapped by the bilayer cloak, the isothermal surfaces and thermal flux lines outside the cloak restore exactly without distortion as if there was nothing.

Figs. 2(b) and 2(c) show the temperature profiles of the perturbation with a single layer of alloy and polystyrene respectively, in which the single layers have the same thickness as the bilayer cloak of 4.2 mm. It is clear that the isothermal surfaces and thermal flux lines are significantly distorted in the both cases. In Fig. 2(c), though the central region is protected similar to the bilayer cloak (thermal flux goes around the central region), external field is severely distorted. Analogy to wavedynamic cloak[5-25], an ideal thermal cloak has to satisfy two conditions: **1)** the external field should be repelled to touch the cloaking region; **2)** the external field outside the cloak should be not disturbed as if there was nothing. Our bilayer thermal cloak completely fulfills the second condition, which has been validated in Fig. 2(d). However, due to ideal thermal insulation material ($\kappa = 0$) is unavailable, thermal energy will slightly diffuse into the cloaking region and raise the temperature as the time elapses. Hence, thermal protection (the first condition) works only transiently, which has also been verified in transformation-based cloak[27,28].

## Measurement Results

For the simplicity of implementation, we consider 2D bilayer cloak as shown in Fig. 1(b). We choose *a*=6 mm, *b*=9.5 mm, and *c*=12 mm. The inner and outer layers of the bilayer cloak are polystyrene and alloy with conductivity of 0.03 W/mK and 9.8 W/mK, respectively. For the host background material, we use a thermal conductive sealant block with conductivity of 2.3 W/mK. The bilayer cloak is cast in the host block with dimensions 45 mm (W), 45 mm (D), and 35 mm (H). Local heating on the left side is achieved by a heat source fixed at 60 °C, and the right side is connected to a tank filled with ice water (0 °C). The cross-sectional temperature profile is captured with a Flir i60 infrared camera, as shown in Fig. 1(c). The photo of fabricated bilayer cloak is shown in Fig. 1(d). An aluminum cylinder with radius of 6 mm is placed in the central region as an object that needs to be cloaked.

**Steady State**

Before we present the results of bilayer cloak, we first discuss two comparison cases. The first reference structure for comparison is a bare perturbation (aluminum cylinder) without bilayer clock. The radius of the perturbation is 6 mm with the same size as cloaking region. We have cast the perturbation in host block as described above and maintained the two sides of the block at

60 °C and 0 °C. Simulated temperature profile is shown in Fig. 3(a), in which isothermal lines are also represented with white color in panel. Obviously, the presence of the perturbation completely alters the temperature profile. The high thermal conductivity of the perturbation attracts the heat flux and makes isothermal lines curved outwards both on its left and right. Measured result of the perturbation is shown in Fig. 3(b), which accords very well with simulated result in Fig. 3(a). The distortion of the external thermal profile can be clearly observed.

The second reference structure for comparison is the object covered by a single layer of polystyrene just like that for the cloak; i.e., this reference structure has no outer layer. The simulated result for temperature distribution is shown in Fig. 3(c), which agrees very well with the measured thermal profile of Fig. 3(d). Obviously, though the central region is protected (thermal flux goes around the central region), the low thermal conductivity of the polystyrene repels the heat flux and makes the isothermal lines near the center of the reference structure significantly curved towards the center.

When the perturbation is wrapped by our bilayer cloak, originally curved thermal profile restore exactly without distortion as if nothing was there, as shown in Fig. 3(e). Measured result of the bilayer cloak is demonstrated in Fig. 3(f). Obviously, the external thermal profile is vertical and not distorted as predicted in Fig. 3(e). As analyzed in theoretical part, for an ideal thermal cloak, the only task is to eliminate the external-field distortion since the inner layer is perfect insulation material. However, in practice we have to use an inner layer with finite thermal conductivity, resulting in that thermal energy will slightly diffuse into the cloaking region and raise the temperature as the time elapses. In the steady state, the to-be-protected central region eventually achieves a constant temperature. Therefore, thermal cloak only can provide a temporal heat protection rather than a permanent protection.

**Time-dependent Case**

To examine the real-time performance of bilayer cloak, we measured the temperature distributions of the three cases in Fig. 3 at different times $t$=1, 10, and 60 min after switching on the power of the heater. Because the temperature of the cloaking region in steady state is close to the room temperature when the heat source is fixed at 60 °C, it is difficult to observe the heat protection as the time elapses. Here the heat source is fixed at 80 °C for unambiguously comparison and observation. The measured results for the first reference structure are shown in the left column of Fig. 4. The strong distortion of the external thermal profile can be clearly observed, and the temperature of the object raises quickly as the time elapses. The measured results for the second reference structure are shown in the middle column of Fig. 4. The significantly distortion of the external thermal profile can also be clearly observed, though the temperature of the central region raises slowly as the time elapses. The measured results for bilayer cloak are depicted in the right column of Fig. 4. It is apparent that the cloak successfully fulfills its task in that the central region is colder than its surrounding by slowing down the external heat flux to enter the cloaking region, which can also be found by comparing the left column and right column of Fig. 4. Furthermore, both of front and rear temperature fronts outside the cloak are always kept nearly planar as the time elapses. This transient behavior once again proves the nearly perfect performance of our bilayer thermal cloak.

Simulated time-dependent temperature distributions for the three cases of Fig. 4 are given in supplementary Fig. S1. The geometric parameters and material parameters in simulations completely accord with those in experiments. Obviously, our experiments in Fig. 4 agree very well with the simulations shown in Fig. S1. For continuous real-time observation, simulated animation movies are given in supplementary Movies S1, S2, and S3, corresponding to Figs. S1(a), (b), and (c), respectively.

It is noted that the temperature of the cloaking region is raised as the time elapses due to the finite heat conductivity of the inner layer. Therefore, it is valuable for engineering application to quantitatively examine the temperature change of the cloaking region over time. We choose (-6, 0) mm and (-6, 13) mm as observation points, which correspond to cloaking region (inside the cloak) and external region (outside the cloak), respectively. For a pure background without any perturbation, the two points should have the same temperature because they are located on the same isothermal line. Fig. 5 shows measured time-dependent temperature curves for the two test points. It is clear that the temperature of the cloaking region is always cooler than external region even after a long time (1.5 hours). Especially in the beginning 10 min, the temperature outside the cloak rises much faster than the inside. In the long-time limit, steady state is achieved, which means the temperature of the cloaking region is constant.

**Conclusions**

In summary, we have experimentally demonstrated an advanced bilayer thermal cloak with naturally available materials. Our design scheme, directly from conduction equation, does not rely on transformation optics[3,4], thus can avoid the problems present in previous cloaking proposals, such as extreme parameters (inhomogeneous, anisotropic, and singular) and complicated fabrication[27-29]. Also, this exact scheme makes sure that the scattering fields are zero, which is different from the plasmonics-based scattering-cancellation technique that only makes the dominant orders zero[2]. Finally, nearly perfect performance can be achieved only employing naturally (commercially) available materials indicating that our advanced scheme may be readily for engineering application. It should be pointed out that the presented work has put additional dimension to the emerging field of phononics: controlling and manipulating heat flow with phonons[31].

**Acknowledgements**

C.W.Q. acknowledges the Grant R-263-000-A23-232 administered by National University of Singapore. T.C.H. also acknowledges the support from the Southwest University (SWU112035).

**Figure captions**

**Figure 1 |** (a) Schematic illustration of 3D bilayer thermal cloak with naturally available materials. (b) Experimental setup for the 2D bilayer thermal cloak. The hot side and cold side of the block are connected to a heater and ice water, respectively. (c) Measurement system for capturing the cross-sectional temperature profile. (d) The photo of fabricated bilayer cloak.

**Figure 2 | Simulated temperature distributions for a 3D bilayer cloak with $a$=6 mm, $b$=9 mm, $c$=10.2 mm.** (a) A bare perturbation (object) with radius of 6 mm. (b) The object is covered by a single layer of alloy with thickness of 4.2 mm. (c) The object is covered by a single layer of polystyrene with thickness of 4.2 mm. (d) The object is wrapped by the proposed bilayer cloak. Streamlines of thermal flux are also represented with white color in panel.

**Figure 3 | Simulated and measured temperature distributions for steady state.** (a) Simulation result of the first reference structure. (b) Measured thermal profile of the first reference structure. (c) Simulation result of the second reference structure. (d) Measured thermal profile of the second reference structure. (e) Simulation result of the proposed bilayer cloak with $a$=6 mm, $b$=9.5 mm, and $c$=12 mm. (f) Measured thermal profile of the bilayer cloak. Isothermal lines are also represented with white color in panel.

**Figure 4 | Measured time-dependent temperature distributions at different times $t$=1, 10, 60 min.** (a) Results for the first reference structure of Fig. 3(b). (b) Results for the second reference structure of Fig. 3(d). (c) Results for the bilayer cloak of Fig. 3(f). Obviously, both of bilayer cloak and insulation coating can make the object colder than the bare object. However, only bilayer cloak can restore isothermal lines exactly without distortion as if there was nothing.

**Figure 5 | Measured time-dependent temperature for test points at (-6, 0) mm and (-6, 13) mm.** Points (-6, 0) and (-6, 13) denote the cloaking region and the surrounding (external region), respectively. The inset shows the temperature difference between the two test points. It is apparent that the cloaking region is always colder than its surrounding even after 1.5 hours.

**Figures**

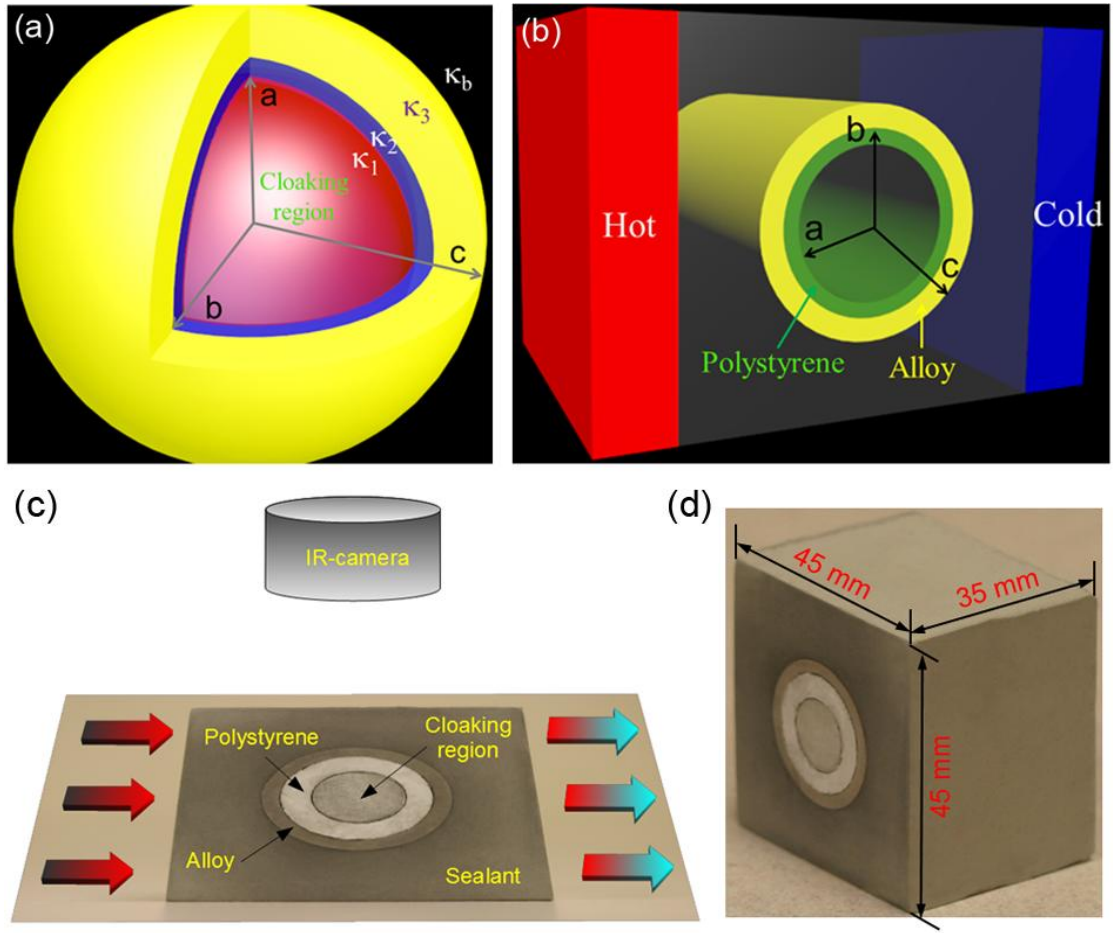

Figure 1

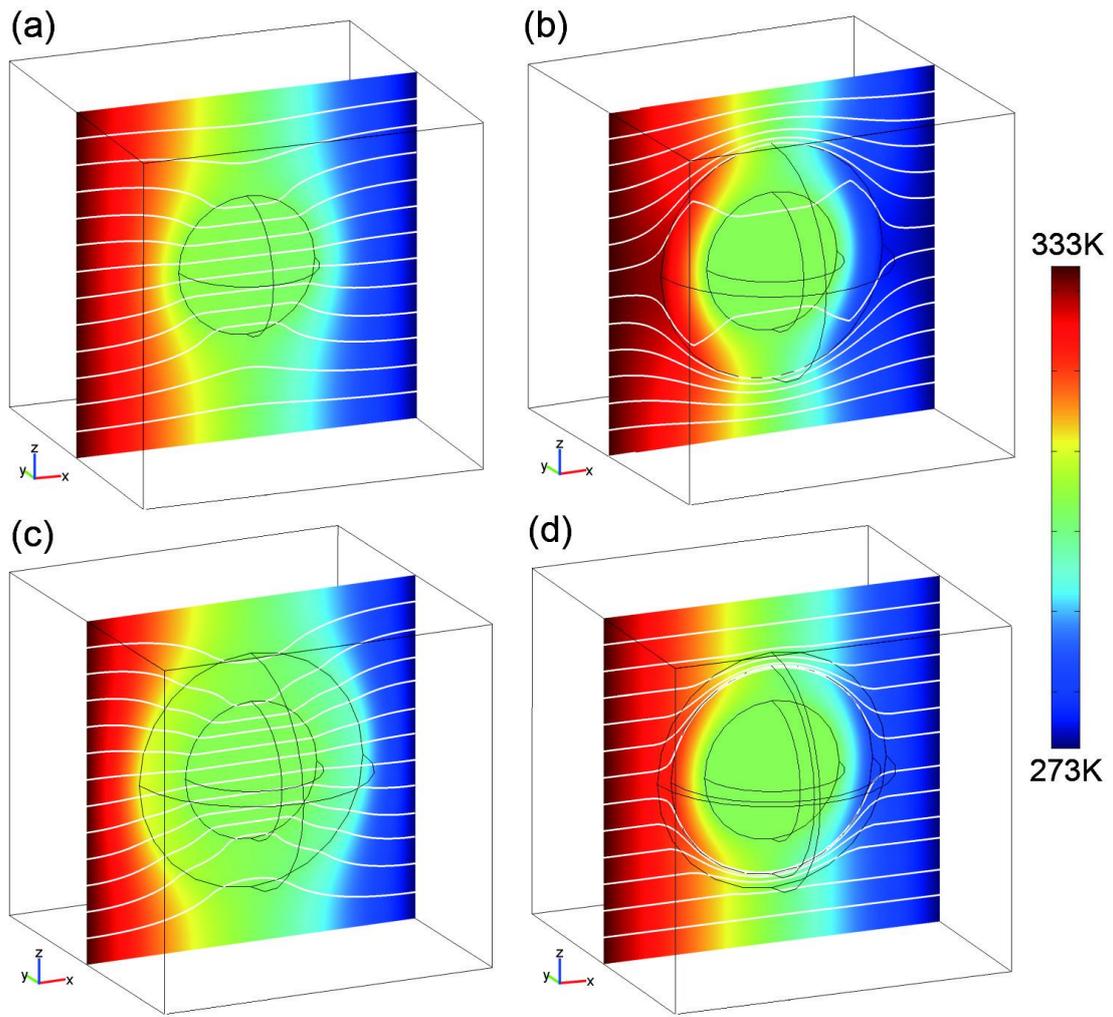

Figure 2

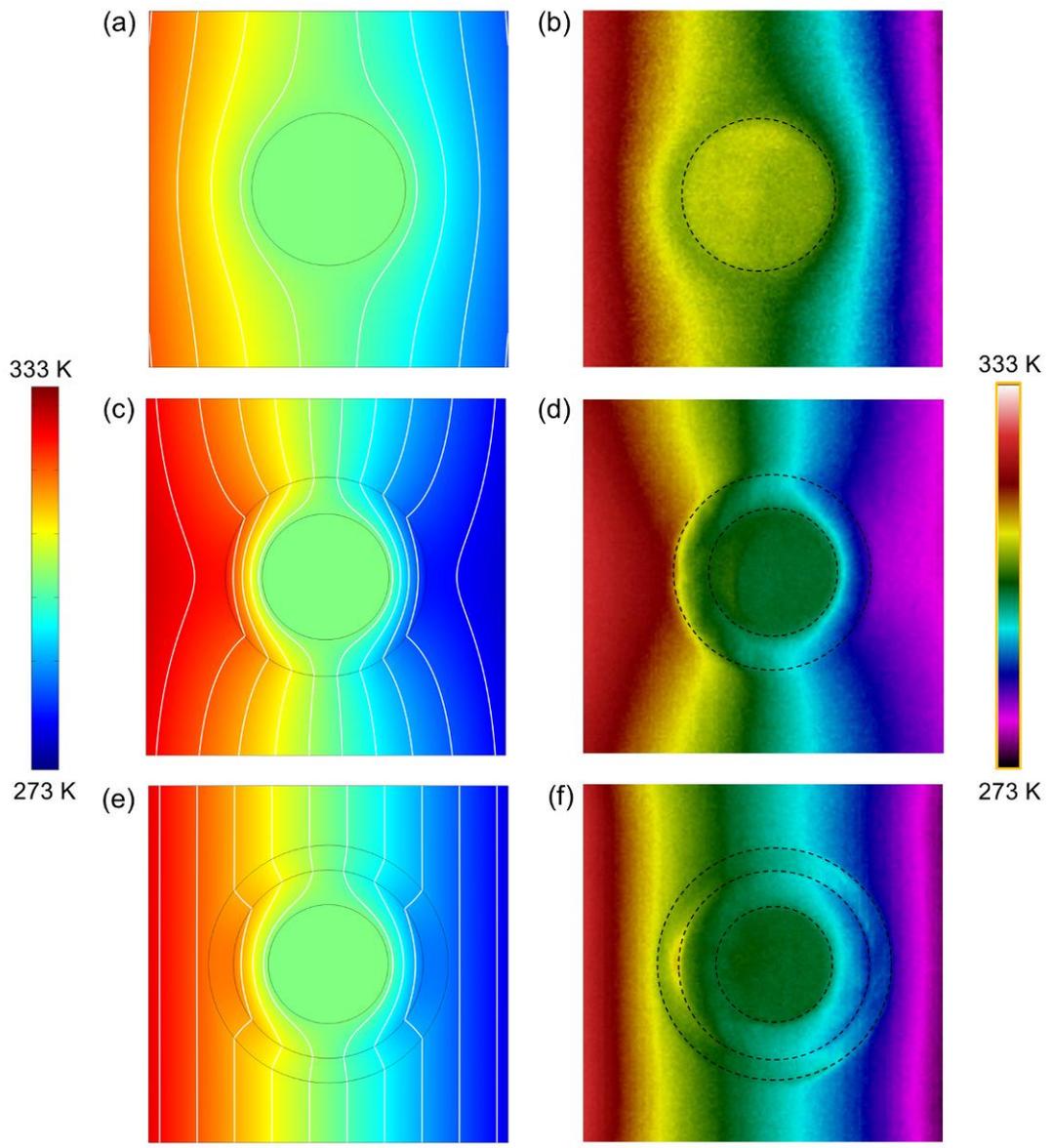

Figure 3

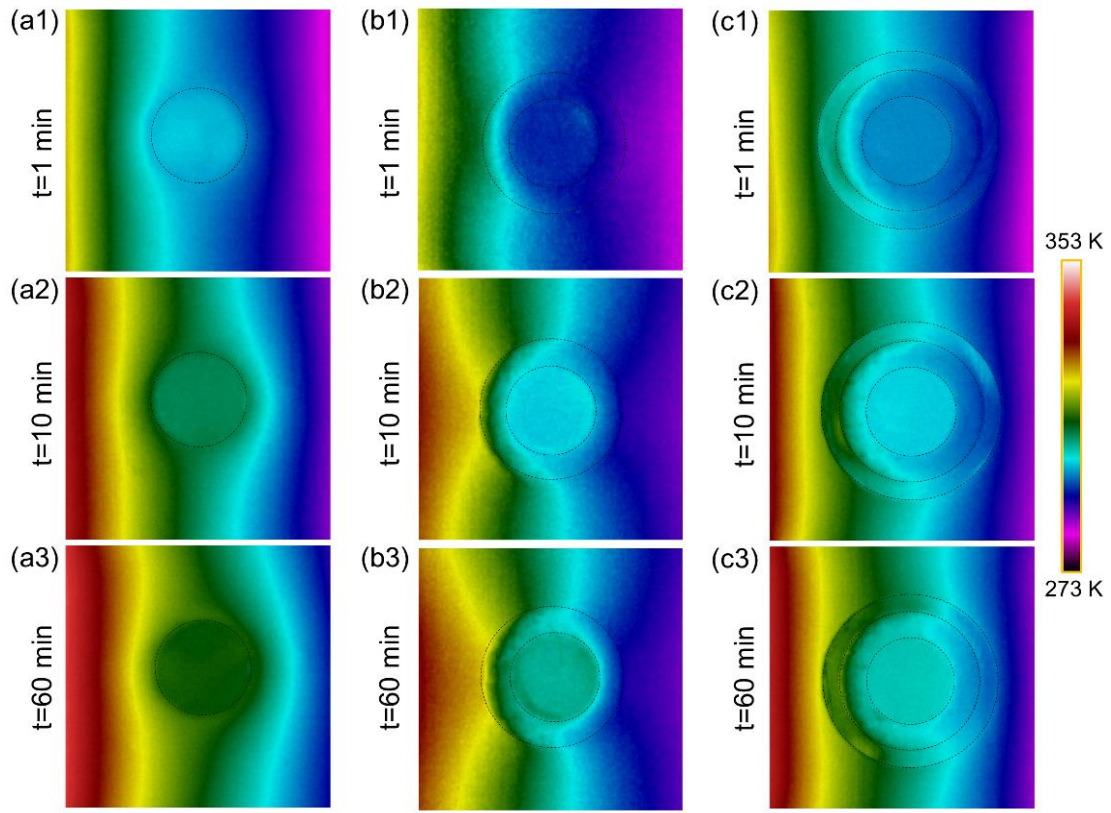

Figure 4

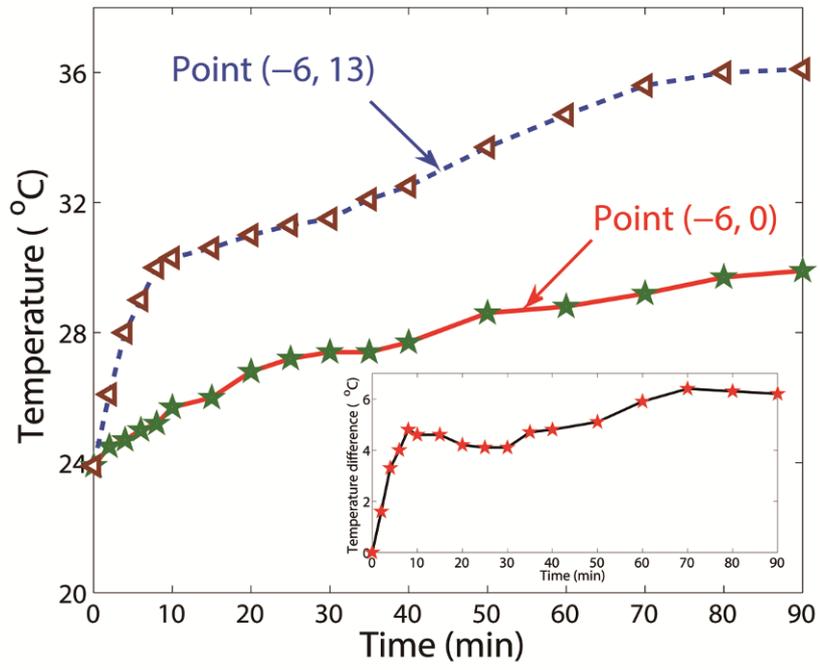

Figure 5

**Supplementary information**

# Bilayer Isotropic Thermal Cloak

Tiancheng Han[1*], Xue Bai[1,2*], John T. L. Thong[1], Baowen Li[2], and Cheng-Wei Qiu[1#]

[1]Department of Electrical and Computer Engineering, National University of Singapore, Kent Ridge 119620, Republic of Singapore.   [#]E-mail: eleqc@nus.edu.sg

[2]Department of Physics and Centre for Computational Science and Engineering, National University of Singapore, Singapore 117546, Republic of Singapore.

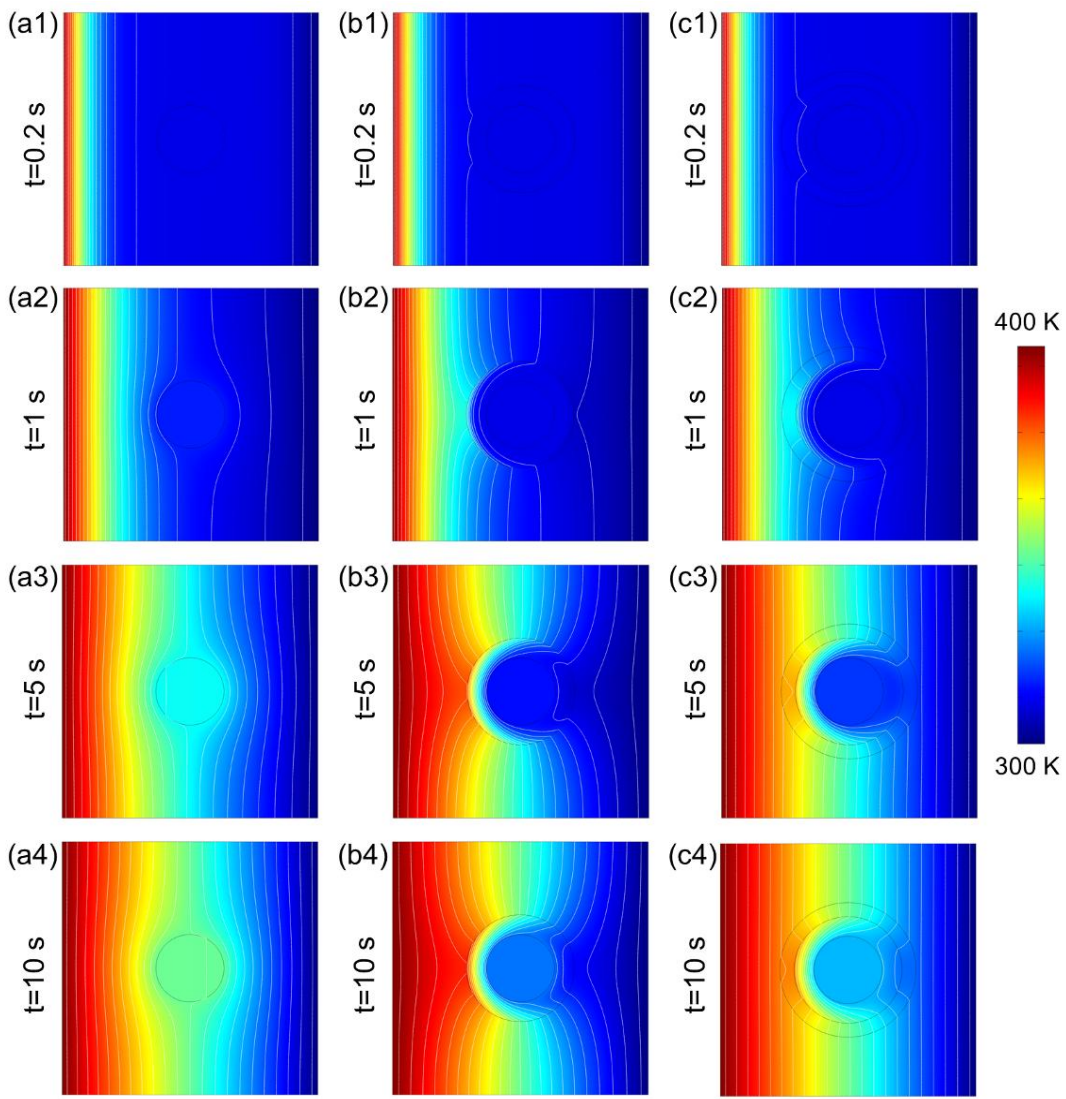

**Fig. S1** Simulated temperature distributions at different times *t* as indicated. The first to third columns correspond to a bare perturbation, the perturbation with a single layer of insulation, and bilayer cloak, respectively. The geometric parameters and material parameters in simulations completely accord with those in experiments.